\documentclass[12pt,preprint]{aastex}

\shorttitle{Cosmological Constraints.}
\shortauthors{Melchiorri et al.}

\begin{document}

\title{Cosmological Constraints from a Combined Analysis
of the Cluster Mass Function and Microwave Background Anisotropies.}

\author{Alessandro Melchiorri\altaffilmark{1},
Paul Bode\altaffilmark{2},
Neta A. Bahcall\altaffilmark{2},
and
Joseph Silk\altaffilmark{1}
}

\altaffiltext{1}{Denys Wilkinson Building, Astrophysics, Keble Road, 
University of Oxford, Oxford, OX1 3RH, UK }
\altaffiltext{2}{Princeton University Observatory, Princeton, NJ 08544}



\begin{abstract}
We present constraints on several cosmological parameters 
from a combined analysis of the most recent Cosmic Microwave Background 
anisotropy data and the Sloan Digital Sky Survey cluster mass function.
We find that the combination of the
two data sets breaks several degeneracies among the parameters
and provides the following
constraints: $\sigma_8=0.76\pm0.09$, $\Omega_m=0.26^{+0.06}_{-0.07}$, 
$h=0.66^{+0.05}_{-0.06}$, $n=0.96 \pm 0.05$,
$\tau_c=0.07^{+0.07}_{-0.05}$.

\end{abstract}




\section{Introduction}

The last few years have seen a spectacular increase in the amount
and quality of available cosmological data.
The new results on the Cosmic Microwave Background
angular power spectrum 
\citep{netterfield, halverson, lee, pearson, scott, benoit} 
have confirmed the theoretical prediction
of acoustic oscillations in the primeval plasma and constrained
theories of large-scale structure formation
\citep[see e.g.][]{wang}.
At the same time, early data from the 2dFGRS \citep{2dfgrs}
and the Sloan Digital Sky Survey \citep{yor00, sto02}
galaxy redshift surveys have provided an unprecedented view 
of the large-scale structure of the universe as traced by galaxies.

Combined analysis of these independent CMB and galaxy data sets
are placing strong constraints on some of the 
fundamental cosmological parameters
\citep{bah99,efstathiou, lahav, mesilk}.
Together with the high-redshift supernovae results 
\citep{perlmutter, filippenko},  a concordance model of a flat,
low-density Cold Dark Matter cosmology has become the current paradigm.

The goal of these analyses is to determine the precise values
of the cosmological parameters of the $\Lambda$-CDM model.
Due to 'cosmic degeneracy', the CMB data 
alone are unable to place tight constraints on several fundamental
parameters, such as the r.m.s. amplitude of the mass fluctuations 
$\sigma_8$, the Hubble parameter $h$ and the optical depth
of the universe $\tau_c$,
even if one restricts the analysis to a flat universe.

In the present {\it Letter} we combine the most recent
CMB anisotropies data with the constraints obtained from the
mass function of clusters of galaxies determined from 
early commissioning imaging data of the SDSS \citep{SDSSmf03}
to break the degeneracy among the cosmological models,
allowing a determination of the best-fit values for individual
parameters.

\section{Method}
\subsection{CMB data and analysis}

We consider a template of flat, adiabatic,
$\Lambda$-CDM CMB spectra computed with CMBFAST
\citep{sz}, sampling the various parameters as follows:
the physical density in cold dark matter 
$\Omega_{cdm}h^2\equiv \omega_{cdm}= 0.05,...0.40$, in steps of  $0.02$;
the physical density in baryons 
$\Omega_{b}h^2\equiv\omega_{b} = 0.009, ...,0.030$,
in steps of $0.003$; 
and the cosmological constant 
$\Omega_{\Lambda}=0.5, ..., 0.95$, in steps of  $0.05$.
For each set of these parameters,
the scalar spectral index $n$ is varied over
the relevant inflationary values of
$n=0.8,..., 1.2$, in steps of $0.02$.
The value of the Hubble constant is not an independent parameter, 
since $h=\sqrt{{(\omega_{cdm}+\omega_b)} / {(1-\Omega_{\Lambda})}}$;
we include a top-hat prior $h=0.7\pm0.2$ \citep{FMGetal01}.
Only models with age $t_0>11$ Gyrs are considered.

We allow for a reionization of the intergalactic medium by
varying the Compton optical depth parameter
$\tau_c$ in the range $\tau_c=0.0,..., 0.45$ in steps of $0.05$.
High values of $\tau_c$ are in disagreement with recent estimates of
the redshift of reionization $z_{re}\sim 6 \pm 1$
\citep[see e.g.][]{Fanetal01,gnedin}, 
which point towards $\tau_c \sim 0.05-0.10$.
However, since the mechanism of reionization is still not clear,
we allow this parameter to vary freely within the above conservative range.
As shown below, the combination of the CMB and CMF data 
provides an independent constraint on this parameter.

For the CMB data, we use the recent results from the
BOOMERanG-98 \citep{netterfield}, 
DASI \citep{halverson}, MAXIMA-1 \citep{lee}, 
CBI \citep{pearson}, VSA \citep{scott} and Archeops 
\citep{benoit} experiments.

The power spectra from these experiments were estimated in
$19$, $9$, $13$, $14$, $10$ and $16$ bins respectively.
For the CBI, we use the data from the MOSAIC configuration \citep{pearson},
spanning the range $2 \le \ell \le 1500$.
We also use the COBE data from the RADPACK compilation \citep{radpack}.

For the Archeops, CBI, DASI, MAXIMA-I and VSA experiments
we use the publicly available correlation matrices and window functions.
For the BOOMERanG experiment we assign a flat interpolation
for the spectrum in each bin $\ell(\ell+1)C_{\ell}/2\pi=C_B$,
and we approximate the signal $C_B$ inside
the bin to be a Gaussian variable.
The CMB likelihood for a given theoretical model is defined by
 $-2{\rm ln} {\cal L}^{CMB}=(C_B^{th}-C_B^{ex})M_{BB'}(C_{B'}^{th}-C_{B'}^{ex})$
where  $M_{BB'}$ is the Gaussian curvature of the likelihood
matrix at the peak.

We consider $7 \%$, $10 \%$, $4 \%$, $5 \%$, $3.5 \%$  and $5 \%$
Gaussian distributed calibration errors (in $\Delta T$)
for the Archeops, BOOMERanG-98, DASI, MAXIMA-1, VSA,
and CBI experiments respectively and we include the beam uncertainties
by the analytical marginalization method presented in \citet{bridle}.
Finally, we rescale the spectrum by a prefactor $C_{10}$, assumed to 
be a free parameter, in units of $C_{10}^{COBE}$.

In order to constrain a parameter $x$ we marginalize over the values of 
the other parameters $\vec{y}$. This yields the marginalized
likelihood distribution
 
\begin{equation}
\mathcal{L}(x) \equiv
        P(x|\vec{\cal C}_B) =
        \int {\cal L}(x,\vec{y})  d\vec{y}.
\label{likelihood1}
\end{equation}

\noindent The central values and $1\sigma$ limits are then found from the 16\%,
50\% and 84\% integrals of $\mathcal{L}(x)$.  


\subsection{Cluster Mass Function Analysis}

We use the cluster mass function (CMF) obtained from the early SDSS 
commissioning data \citep{SDSSmf03}.
This CMF was derived from 294 clusters 
in the redshift range z = 0.1 - 0.2 selected by the 
Hybrid Matched Filter (HMF) method. 
The HMF mass function was compared with large scale cosmological simulations 
as well as with model predictions, as discussed in \cite{SDSSmf03}. 
If the Hubble constant and spectral index are kept constant at
$h$=0.72 and $n$=1, then the best-fit relation between amplitude
and density can be summarized as $\sigma_8 \Omega_m^{0.6} = 0.33$.  
The 68, 95, and 99\% confidence contours (allowing $h$ and $n$ to vary
as in the previous section) are shown by the dashed curves in Figure 1.

The shape of the cluster mass function only partially breaks 
the degeneracy between 
$\Omega_m$ and $\sigma_8$ in the above relation:
low values of $\sigma_8$ yield a steeper CMF shape at the high-mass end (i.e.,
fewer high mass clusters) than do low $\Omega_m$ values (which produce 
a flatter CMF shape). 
The cluster mass function prefers a low value for the mass density
parameter and a relatively high value for the amplitude $\sigma_8$:
the best-fit parameters for
the HMF clusters are $\Omega_m$=0.18 and $\sigma_8$=0.92; similar
results are obtained for SDSS clusters selected by the the maxBCG method
\citep[see][]{SDSSmf03}.
The above $\sigma_8-\Omega_m$ relation is consistent with recent results 
from the X-ray cluster temperature function \citep[see][]{uros},
and from cosmic shear lensing observations 
\citep{ike02, rei02, hamana02, jarvis02}. 
It also implies that for a mass density of $\Omega_m$=0.3 the 
relevant amplitude is $\sigma_8$ = 0.7.

The observed CMB spectrum of fluctuations suggests a lower amplitude value
for $\sigma_8$ (which is, however, degenerate with the optical depth parameter
for CMB), and a somewhat larger value for the mass density parameter 
($\Omega_m \sim 0.3$ for $h$=0.72), than 
given by the cluster mass function above 
\citep[see e.g.][and references therein]{lahav, mesilk}.
However, the CMB and the cluster mass function results are 
consistent with each other within one sigma. Combining the CMB and CMF data
will clearly result in intermediate values for the
cosmological parameters, shifting the CMF constraints 
presented above towards a somewhat lower amplitude
and higher mass density regime.

We combine the cluster mass function results with
those of the CMB by multiplying the two likelihoods
${\cal L}^{CMB}{\cal L}^{CMF}$, using the same range of parameters 
($\Omega_m$, $\sigma_8$, $h$, $n$) and marginalizing
over the nuisance parameters as discussed in the previous section.

\section{Results}

The main result of our analysis is presented in Figure 1, where we
plot likelihood contours in the $\Omega_m-\sigma_8$ plane
for the two datasets, separately and combined.
It can be seen that both the CMB and CMF datasets are affected by
degeneracies between $\sigma_8$ and $\Omega_m$. 
In the case of the CMF, an increase of $\Omega_m$ results in a 
larger number of clusters, and $\sigma_8$ must be reduced
to bring the predicted CMF back in line with observations.
On the other hand, the CMB is only weakly sensitive to the tradeoff
between density and amplitude, so models 
with higher $\Omega_m$ and $\sigma_8$ can be in
agreement with CMB data. The degeneracy in the CMB dataset
is therefore opposite to the one in the CMF data, and the
two measurements complement each other.
The combination of the two
datasets provides the constraints: $\Omega_m=0.26_{-0.07}^{+0.06}$ and
$\sigma_8=0.76\pm 0.09$ at $1 \sigma$ confidence level,
as shown in Figure 1.

The CMB+CMF combination can also break additional degeneracies.
The optical depth $\tau_c$ and the
Hubble parameter $h$ are better determined after the
inclusion of the CMF data.
From the combined analysis, we obtain (at $1-\sigma$ C.L.)
$h=0.66_{-0.06}^{+0.05}$, $n=0.96\pm0.05$ and $\tau_c=0.07_{-0.05}^{+0.07}$;
these values can be compared with $h=0.70_{-0.13}^{+0.11}$,
$n=0.98_{-0.07}^{+0.09}$ and $\tau_c < 0.29$ from the CMB-only 
analysis.

\section{Conclusions}

We combine the Cosmic Microwave Background anisotropy
and SDSS cluster mass function data to produce 
constraints on several cosmological parameters.
The complementary nature of the two data sets
breaks existing degeneracies among cosmological parameters.
The CMB data tend to indicate a higher $\Omega_m$ and a lower amplitude 
$\sigma_8$ than suggested by the cluster data; thus the combined result 
for $\Omega_m$ is pulled to a lower value than suggested by the CMB alone, 
and the amplitude $\sigma_8$ is lower than suggested by the cluster 
data alone.
We find the combined data suggest a mass density 
$\Omega_m=0.26_{-0.07}^{+0.06}$; a normalization of the matter power spectrum 
$\sigma_8=0.76\pm0.09$; Hubble parameter $h=0.66^{+0.05}_{-0.06}$;
a nearly scale-invariant spectrum of primordial fluctuations 
$n=0.96 \pm 0.05$; and
an optical depth of the universe $\tau_c=0.07^{+0.07}_{-0.05}$. 
We have restricted the analysis to flat universes; 
if this is the case, the density of mass-energy in the universe
is dominated by a form other than ordinary matter.  

\acknowledgments

We acknowledge a grant from the Oxford-Princeton Research Partnership
Initiative.
AM is supported by PPARC;
PB is supported by the National Computational Science Alliance
under NSF Cooperative Agreement ASC97-40300.

\begin{figure}
\plotone{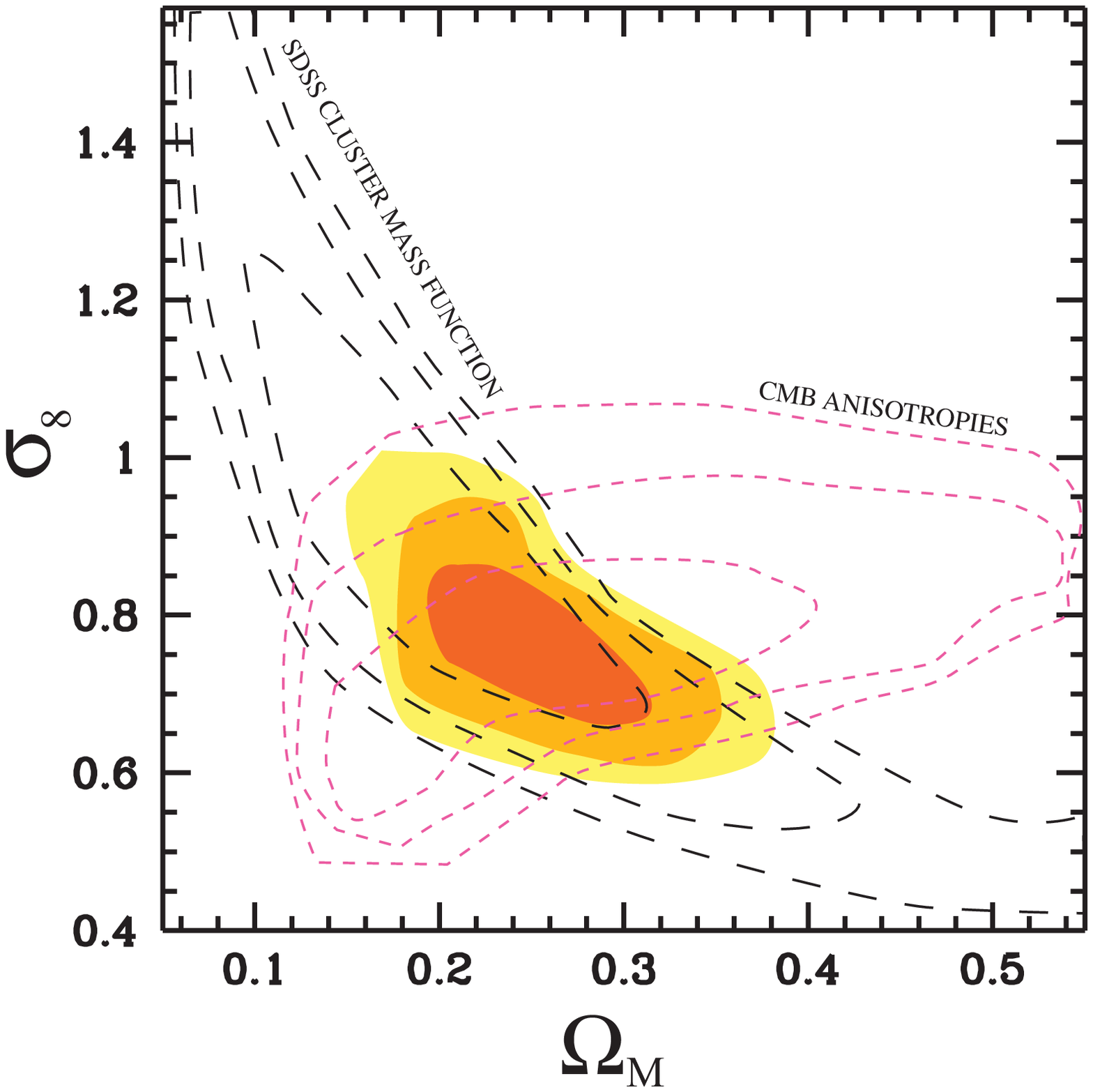}
\caption{Confidence contours in the $\Omega_m-\sigma_8$ plane from
the CMB+CMF combined analysis.  Bold dashed and faint dashed curves
represent the 68, 95, and 99\% confidence contours from the CMF
and CMB data, respectively.  The shaded regions represent the 
corresponding ranges allowed by the combined analysis.}
\label{fig1}
\end{figure}

\end{document}